\def\hone{\ifmmode{\rm H \/ {\sc i}}\else{H\/{\sc i}}\fi} 
\def\lesssim{\mathrel{\hbox{\rlap{\hbox{\lower4pt\hbox{$\sim$}}}\hbox{$<$}}}}
\def\gtrsim{\mathrel{\hbox{\rlap{\hbox{\lower4pt\hbox{$\sim$}}}\hbox{$>$}}}}

\documentclass[useAMS,usenatbib]{mn2e}
\usepackage{epsfig}
\usepackage{float}
\usepackage{afterpage}

\newcommand{\apj}{{ ApJ}}
\newcommand{\aj}{{ AJ}}
\newcommand{\apjl}{{ ApJL}}
\newcommand{\apjs}{{ApJS}}

\newcommand{\mnras}{{MNRAS}}
\newcommand{\araa}{{ ARA\&A }}
\newcommand{\aap}{{ A\&A}}
\newcommand{\aaps}{{ A\&A Supp}}

\newcommand{\nat}{{ Nature}}
\title
[Kinematical structure of arcs]
{The kinematical structure of gravitationally lensed arcs}
\author
[Ole M\"oller \& Edo Noordermeer]
{
Ole M\"oller$^{1,2}$\thanks{E-mail: ole@mpa-garching.mpg.de}, Edo Noordermeer$^1$\\
$^1$Kapteyn Astronomical Institute, University of Groningen, PO Box 800, 9700 AV Groningen, the Netherlands.\\
$^2$Max-Planck-Institut f\"ur Astrophysik, Karl-Schwarzschild-Strasse
1, D-85741 Garching, Germany.\\
$^\star$E-mail: ole@mpa-garching.mpg.de
}
\begin{document}

\pubyear{200?}
\maketitle

\begin{abstract}
In  this paper  the expected properties  of the  velocity fields  of strongly
lensed  arcs behind  galaxy  clusters are  investigated. The  velocity
profile along  typical lensed arcs is  determined by ray tracing light 
rays from a model source galaxy through parametric cluster toy-models consisting  of  individual  galaxies embedded  in  a  dark
cluster  halo. We  find that  strongly  lensed arcs  of high  redshift
galaxies show  complex velocity structures  that are sensitive  to the
details  of the  mass distribution  within the  cluster, in particular at small
scales. From fits to the simulated imaging and kinematic data we demonstrate that
reconstruction  of  the  source   velocity  field  is  in  principle
feasible: two dimensional kinematic information obtained with Integral
Field  Units (IFU's) on  large  ground based  telescopes  in combination  
with adaptive optics  will allow the  reconstruction of rotation  curves of lensed 
high redshift galaxies.  This makes it possible to  determine the mass-to-light 
ratios  of galaxies at redshifts $z>~1$ out to  about  2-3 scale
lengths with  better than  $\sim 20\%$ accuracy.  We also  discuss the
possibilities of using two dimensional kinematic information along the
arcs to give additional constraints on the cluster lens mass models.
\end{abstract}
\section{Introduction}
In recent years, the high redshift universe has become a focus of attention
for observational cosmology as a testbed for theoretical models of galaxy
formation and cosmology. The detection of high redshift galaxies and quasars
has led to the first observational tests of reionisation models
\citep{haiman2003,ciardi2005} and high redshift absorption line systems have helped to constrain spin temperatures and the fraction of cold neutral gas in high redshift galaxies \citep{kanekar2003}. Source counts in the sub-mm and the infrared have constrained the star formation at high redshifts \citep{hughes1998,blain1999c}. However, despite these tremendous advances in the field, a few questions pertaining to the high redshift universe and the evolution of galaxies remain difficult to address. Foremost, there is the open question as to the role of the dark matter in galaxy evolution. Even in the local universe, the presence of dark matter can only be inferred indirectly through its gravitational effect; for example by measuring the rotation curves of galaxies. 

Rotation curves have now been measured for a large sample of nearby
galaxies, both in optical wavelengths \citep{mathewson1992,persic1995,palunas2000}
and using the 21cm emission line of neutral hydrogen
\citep{verheijen2001,swaters1999}. They have improved our knowledge of
the systematics of dark matter in nearby galaxies, but they have also
led to a number of new questions that still need to be addressed in 
theoretical models of galaxy formation. The strong dependence of
rotation curve shape and amplitude on the optical luminosity indicates
a tight coupling between luminous and dark matter that is not expected
from the current models \citep{persic1996,donato2004}. There are
also indications from the rotation curves of low surface brightness
galaxies that galactic dark matter haloes contain a constant density
core, as opposed to the $r^{-1}$ cusp predicted by simulations
\citep{deblok2001,deblok2002}. 

Using $H\alpha$ emission, rotation curves of galaxies have been measured
accurately up to redshifts of about $0.3$. $H\alpha$ emission has been
detected out to much larger distances and has been used to constrain the star
formation rates in galaxies at redshifts greater than 2
\citep{bunker1999}. Attempts have also been made to measure the rotation curves of galaxies up to
redshifts of $z\sim1$ \citep{vogt1996,vogt1997,hudson1998,boehm2004}, but at these redshifts it is
difficult to determine the inclination angle of the rotating disc accurately
and to obtain sufficient spatial resolution. So far the limited spatial resolution has
only allowed determinations of the total velocity dispersion rather than actual rotation
curves: therefore most authors concentrated on constraining the evolution of
the Tully-Fisher
relation \citep{tully1977}. 

Galaxies that are at a higher redshift but are strongly lensed may well be magnified by factors of 10 or more \citep[e.g.][]{blandford1992}. Exploiting this effect, gravitational lensing has already been successfully used as a tool to study high redshift sources in greater detail than would be otherwise possible \citep{pello1999,ellis2001,richard2003}.
Lensed galaxies may have a large enough area and luminosity to make a measurement of their rotation curve possible, despite their larger distance.  \citet{narasimha1993} demonstrated that it is in principle possible to measure the kinematics of galaxies that are much more distant in this way.
Since integral field spectrometers had not been developed at the time of the
publication of that paper, it focused on measuring the velocities along
straight arcs. Long slit spectroscopy of straight arcs has been carried out
successfully by \citet{bunker2000} and \citet{mehlert2001}. Recently,
improvements in resolution and sensitivity of integral field spectrometers have made it
possible to obtain kinematics of moderately magnified background galaxies \citep{swinbank2003}. 

At intermediate redshifts between $z\sim0.3-1$, gravitational lensing also provides
constraints on the dark matter content of the inner $10-200\,\mathrm{kpc}$ of lens galaxies \citep[e.g.][]{koopmans1998,koopmans2003} and lens clusters \citep[e.g.][]{smail1995,mellier1999,kneib2003}. Determinations of
the dark matter distribution using gravitational lensing rely on accurate data to provide sufficient constraints on the lens mass
model. Degeneracies in the lens model usually exist and may have important consequences
for determinations of cosmological parameters and mass density
profiles from lensing \citep{williams1999,zhao2003,meneghetti2004,dalal2004}. Additional
constraints help to break such degeneracies. For galaxy lenses, spectroscopic studies have been used to break degeneracies in the lens models and constrain their mass distribution further \citep{koopmans2002a}. 

In this paper we investigate the possibilities of measuring rotation
curves of lensed galaxies with current and upcoming instruments and discuss 
the additional constraints on the foreground lens mass model that may be obtained from two dimensional spectroscopic data. 

We begin by describing the theoretical framework, the ray-tracing method and the
source and cluster lens models in
\S\,\ref{datacubes.sec}. In \S\,\ref{knownarcs.subsec} we briefly
discuss the optical properties of known lensed arcs. In
\S\,\ref{lensed.sec} we present the kinematics of
simulated arcs and discuss their generic properties. In 
\S\,\ref{reconstruct.sec} we discuss how the kinematic profile of
the source can be reconstructed from lens modeling and discuss their
dependence on the lens mass model. We address the observational
possibilities using current and future instruments -- in particular
the use of Integral Field Units (IFUs) -- in
\S\,\ref{observations.sec}. We conclude with a discussion and summary in \S\,\ref{conclusions.sec}.

Throughout this paper we use a standard $\Lambda$CDM cosmology
with $\Omega_{\Lambda}=0.7$, $\Omega_{\mathrm{m}}=0.3$ and
$H_0=70\,\mathrm{km\,s^{-1}\,Mpc^{-1}}$.

\section{Simulating data cubes of lensed high redshift galaxies}
\label{datacubes.sec}

\subsection{Lensing theory}
\label{theory.subsec}
A background source at a redshift $z_{\mathrm{s}}$ that is located at a position $\mathbf{\beta}$ will appear at a position $\mathbf{\theta}$ on the sky if it is lensed by a massive foreground object at redshift $z_{\mathrm{l}}$ so that
\begin{equation}
\mathbf{\theta}=\mathbf{\beta}+\frac{D_{\mathrm{LS}}}{D_{\mathrm{OS}}}\alpha(\mathbf{\theta}),
\label{lensing.equation}
\end{equation}
where $D_{\mathrm{LS}}$ and $D_{\mathrm{OS}}$ are the angular diameter distances from lens to source and from observer to source, respectively. The deflection angle $\mathbf{\alpha}$ at position $\mathbf{\theta}$ on the lens plane is given by
\begin{equation}
\mathbf{\alpha}(\mathbf{\theta})=\frac{D_{\mathrm{OS}}}{\pi D_{\mathrm{LS}}}\int\kappa(\mathbf{\theta'})\frac{\mathbf{\theta}-\mathbf{\theta'}}{\left|\mathbf{\theta}-\mathbf{\theta'}\right|^2}\,d^2\theta',
\label{deflection.equation}
\end{equation}
where $\kappa$ is a dimensionless quantity related to the surface mass density $\Sigma$ through,
\begin{equation}
\kappa(\mathbf{\theta})=\Sigma(\mathbf{\theta})\times\frac{4\pi GD_{\mathrm{OL}}D_{\mathrm{LS}}}{c^2D_{\mathrm{OS}}}.
\end{equation}

The magnification of an image at position $\theta$ of a point source at position $\beta$ is given by
\begin{equation}
\mu(\mathbf{\theta})=\frac{1}{[1-\kappa(\mathbf{\theta})]^2-\gamma^2(\mathbf{\theta})},
\end{equation}
where $\mathbf{\gamma}=(\gamma_x,\gamma_y)$ is the total shear at the image position, which is
related to the deflection angle, and hence the lensing mass distribution, through:
\begin{eqnarray}
\gamma_x &=& \frac{1}{2}\left(\frac{\partial\alpha_x}{\partial
  x}-\frac{\partial\alpha_y}{\partial y}\right), \\
\gamma_y &=& \frac{\partial\alpha_y}{\partial x}=\frac{\partial\alpha_x}{\partial y}.
\end{eqnarray}
For extended sources, different parts of the source will be magnified by different amounts, leading to differential magnification \citep{blain1999}. Given a magnification map on the source plane $\mu_{\mathrm{s}}(\mathbf{\beta})$ and the source flux $F_{\mathrm{s}}(\mathbf{\beta})$, the observed total flux is
\begin{equation}
 F_{\mathrm{tot}}=\int F_{\mathrm{s}}(\mathbf{\beta})\mu_{\mathrm{s}}(\mathbf{\beta})\,d^2\beta.
\end{equation}

The type of observations we are interested in here, namely spatial spectroscopy using integral field units, are characterised by several channels at frequencies $\nu$, of bandwidth $\Delta\nu$. The observed flux in an interval between $\nu$ and $\nu+d\nu$ would be
\begin{equation}
F(\nu)\,d\nu=\int F_{\mathrm{s}}(\nu,\mathbf{\beta})\mu_{\mathrm{s}}\left(\mathbf{\beta}\right)\,d^2\beta d\nu,
\end{equation}
and so, for finite bandwidths,
\begin{equation}
F(\nu)=\int_{\nu}^{\nu+\Delta\nu}\!\!\!\int F_{\mathrm{s}}\left(\nu,\mathbf{\beta}\right)\mu_{\mathrm{s}}\left(\mathbf{\beta}\right)\,d^2\beta d\nu.
\end{equation}
Thus, if $\mu_{\mathrm{s}}\left(\mathbf{\beta}\right)$ is known from the lens model, it is straightforward to obtain the total observed flux at a given wavelength, given a model of the source. This can be done simply by summing over all the pixels of a map that is the product of the source flux at the observed wavelength and the source magnification map. 

\subsection{Modeling the source and creating mock kinematic data}
\label{modelling.subsec}

In order to create a model data cube for the source, we need to make
assumptions about the gas distribution and its kinematics. 

For the latter, we take the `Universal Rotation Curve' as derived by
\citet{persic1996}. We assume an $L_*$ galaxy, for which their
equations reduce to:  
\begin{equation}
V^*(x) = V_0^* \sqrt{\frac{1.4x^{1.2}}{(x^2 + 0.6)^{1.4}} + 
  \frac{1.1 x^2}{x^2 + 2.2}}, 
\end{equation}
with $x=R/R_{\mathrm{opt}}=R/3.2h$ being the radius expressed in units  
of $R_{\mathrm{opt}}$, which is the radius encompassing $83\%$ of the 
total light; $h$ is the disc scale length. We set the parameter $V_0^* = 201.5 \,
\mathrm{km/s}$. 
Defining the radius expressed in disc scale lengths, $x_{\mathrm{h}}
= R/h$, we get:  
\begin{equation}
V^*(x_{\mathrm{h}}) = V_0^* \sqrt{\frac{0.3 
    x_{\mathrm{h}}^{1.2}}{(0.1 x_{\mathrm{h}}^2 + 0.6)^{1.4}} + 
  \frac{0.1 x_{\mathrm{h}}^2}{0.1 x_{\mathrm{h}}^2 + 2.2}}.
\label{URC.eq}
\end{equation}

In reality, galaxies do not follow the Universal Rotation Curve
exactly, and galaxies with different masses will have different
shapes for their rotation curves. Furthermore, \citet{persic1996}
derived their equations from rotation curves of galaxies in the
local universe, and their results may not be directly applicable
to the high-redshift galaxies we are studying here. However, the exact
shape and amplitude of the rotation curve is not critical for the
study we present here and different assumptions would not affect our
conclusions.   

For the gas distribution, we assume an exponential radial profile:
\begin{equation}
I = I_0 \exp(-x_{\mathrm h}).
\end{equation}

We adopt a radial scale-length of $h = 0.4$\,arcsec,
corresponding to approximately 4 kpc at the assumed 
source redshift of $z_{\mathrm{s}}=1.5$. Furthermore we assume that
the disc of the galaxy is inclined at an angle of 50 degrees with respect to the line of sight. 
The vertical density distribution of the gas is assumed to be
exponential as well, with scale-height ${\mathrm z_0}$ equal to 1/20th
of the radial 
scale-length $h$. This means we effectively assume the galaxy to be razor
thin, but the effect of a larger vertical scale-height on the simulated
data cube is very small for the moderate inclination of 50 degrees we assume here.  

To create the model data cube, we use the GALMOD task in GIPSY: The Groningen Image Processing SYstem. It
uses a Monte-Carlo integration to fill the model galaxy with small
gas clouds, following the exponential distribution given above. 
For each cloud, the radial velocity is calculated on the basis of its
position in the disc and the rotation velocity at its radius, and each
cloud is assumed to emit a gaussian  emission line with an intrinsic
velocity width of $10\, {\mathrm {km/s}}$. The final data cube consists
of channel-maps 
spaced $\approx 8\, {\mathrm {km/s}}$, with spatial pixels of $0.02
$\,arcsec, or $h/20$. 

\begin{figure}
\epsfig{file=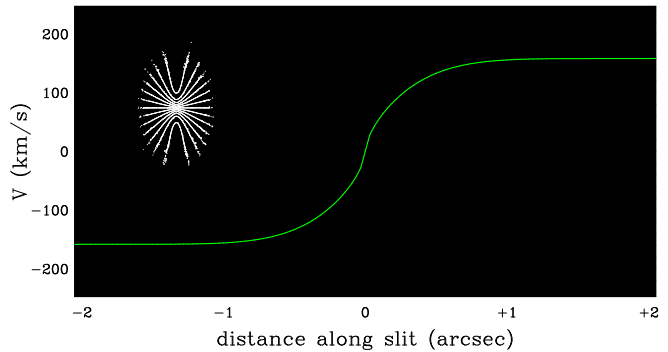,width=4.8cm,angle=-90}\\
\caption{Position-velocity diagram of the source along the major
axis. Contours correspond to the maximum intensities at 1, 3 and 5
radial scale lengths. The green line indicates the rotation curve,
multiplied by $\sin (50^\circ)$.  
The inset in the top left shows the velocity map of the source. The
distances along the axes are in arcseconds. Different colours indicate
radial velocities, ranging from -175 to 175 $\,{\mathrm {km/s}}$, as
indicated by the colour bar; the contours range from -150 to
150$\,{\mathrm {km/s}}$ with intervals of $25\, {\mathrm {km/s}}$.}   
\label{fig.single.vel}
\end{figure}
In Fig.\,\ref{fig.single.vel}, we show a
cut through the data cube along the major axis of the galaxy. 

From the data cube, we derive an image of the integrated gas emission
by adding up, at each pixel position, the signal in the individual
channel maps. The velocity field is derived by fitting gaussians to
the line profiles. It is shown in the inset in Fig.\,\ref{fig.single.vel}, and
resembles those observed in nearby spiral galaxies \citep{verheijen2001,garrido2002}.

In the first instance, we use a very high number of small gas clouds
in the Monte-Carlo integration. This results in a very smooth gas
distribution and velocity field. We can simulate a very patchy gas
distribution by limiting the number of clouds to a very small number.
We will discuss how such strong substructure in the source will affect
our results in \S\,4.2 and \S\,5.5.
%
%
\subsection{Modeling the lensing potential}
\label{lensmodel.subsec}
Simulations of strongly lensed arcs in clusters have been performed by
several                                                          groups
\citep{bartelmann1994,wu1996,hamana1997,bartelmann1998,wambsganss1998a,meneghetti2001,meneghetti2003}. Most
of these studies make use of a cluster model from  N-body simulations. In
contrast,  the basic  approach  in our  work is to  use an analytic, parametric
cluster  model. Analytic
models  have been  used to  predict the  arc statistics  for different
cosmologies by \citet{wu1996}, \citet{cooray1999} and \citet{oguri2001}, finding a
much lower cross section for  the formation of arcs than when clusters
are  modeled  directly from  N-body  simulations.  As  pointed out  by
\citet{meneghetti2003} massive cluster substructure and assymmetry are
important and explain   part   of   the
discrepancy. \citet{wambsganss2004} recently demonstrated that the source redshift is also
an important factor affecting the statistical incidence of strongly lensed
arcs. Furthermore,
since small scale structure may have a very significant, local effect, the
actual structure and shape of the arcs will be influenced strongly by
the presence of nearby mass concentrations. As shown by \citet{meneghetti2000}
and \citet{flores2000} cluster galaxies are unlikely to be massive enough to
affect the arc statistics -- more massive substructures and asymmetries are
needed. However, structures on galaxy scales are very important for detailed
modeling of observed arc systems \citep{kassiola1992,broadhurst2005}. When such structures
lie close to extended arcs, the shape of the arcs may be strongly affected. Therefore, a realistic  parametric cluster  model that  includes the
individual galaxies is used  here, incorporating the substructure that can
crucially affect the appearance of strongly lensed arcs behind cluster
lenses. 

In order to simulate the lensing potential of a cluster, we create 6
realizations of a mock
cluster at redshift $z_{\mathrm{l}}=0.3$ by arranging 70 galaxies randomly
around the centre of a common dark matter halo. The distribution of galaxy
positions follows a gaussian with a standard deviation of
$\mathrm{var}_{pos}=65\,$arcsec around the cluster centre. The angular
distribution is isotropic. Most clusters are expected to have a
moderately elliptical potential and galaxy distribution. Such an ellipticity
introduces additional parameters into our model but does not significantly affect our main
results. We will discuss the effect of lens ellipticity and profile in detail
in \S\,\ref{nfwsect}.  
Each galaxy is modeled as a singular isothermal sphere (SIS) of the form,
\begin{equation}
\Sigma(r)=\frac{\sigma_{\mathrm{v}}}{2Gr}.
\end{equation}
For each galaxy, the    velocity   dispersion   $\sigma_{\mathrm{v}}$   is
determined randomly  from a gaussian  distribution of
mean   $\overline{\sigma}^0_{\mathrm{v}}=130\,\mathrm{km\,s^{-1}}$  and  standard
deviation  $\mathrm{var}^0_{\sigma}=92\,\mathrm{km\,s^{-1}}$. In order  to investigate  the effect  of the
amount of  substructure due to  the cluster galaxies,  we parameterise
the relative contribution of the cluster galaxies to the total cluster
mass, using a single  parameter, $\Gamma$. The mean velocity dispersion for
the                  SIS                  galaxies                  is
$\overline{\sigma}_{\mathrm{v}}=\Gamma\overline{\sigma}^0_{\mathrm{v}}$, where $\overline{\sigma}^0_{\mathrm{v}}=130\,\mathrm{km\,s^{-1}}$.

\begin{figure}
\epsfig{file=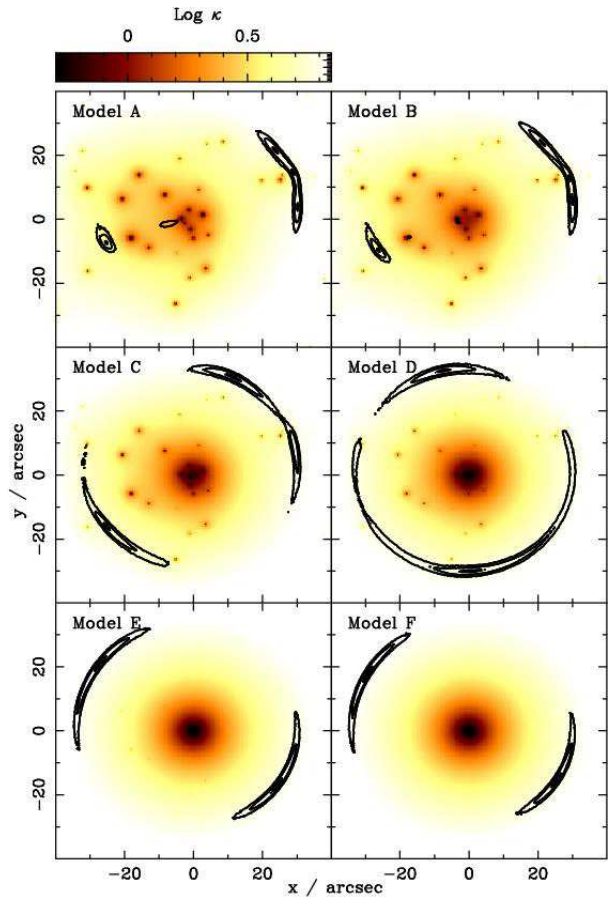,width=8.0cm,angle=180}\\
\caption{The surface mass density and lensed arcs of the model lensing
  clusters. The panels are for models A-F in Table\,1. The colour-scale shows
  the surface mass density of the cluster. The lines show the contours that contain 10, 68
  and 90 per cent of the total source flux. The value of $\Gamma$
decreases from left to right and top to bottom. Note that for larger values of
  $\Gamma$, that is, for larger amounts of substructure, the arcs are broken
  and distorted in several places due to the presence of the individual galaxies.
}
\label{clustermodelsA.fig}
\end{figure}

In addition to the galaxies, we add a cored isothermal sphere (CIS) at the
centre of the cluster, to model both the effect of a common halo and a
central cD galaxy. The surface mass density of the CIS is given by
\begin{equation}
\Sigma(r)=\frac{\sigma_{\mathrm{cl}}^2}{2G}\frac{1}{\sqrt{(r^2+r_{\mathrm{c}}^2)}},
\end{equation}
where $r_{\mathrm{c}}$ is the core radius and
$\sigma_{\mathrm{cl}}$ is the velocity dispersion at $r\to\infty$.
The CIS is centred on the mean galaxy position. Its maximal velocity
dispersion is set to
$\sigma^{\mathrm{max}}_{\mathrm{cl}}=1281\,\mathrm{km\,s^{-1}}$, and it has a core radius of $10$\,arcsec, corresponding
to $110\,\mathrm{kpc}$ at $z=0.3$. For values of $\Gamma>0$, that is, for an increased mass fraction in galaxies, the cluster velocity dispersion is decreased below $\sigma^{\mathrm{max}}_{\mathrm{cl}}$ to keep the total mass inside the Einstein radius constant at $M_{\mathrm{ER}}=1.7\times10^{14}M_{\odot}$.
By varying
the relative  values of $\Gamma$  and the cluster  velocity dispersion
$\sigma_{\mathrm{cl}}$, the degree   of
`clumpiness' in the cluster is changed while keeping the Einstein radius constant at $32$\,arcsec.
\begin{table}
\label{clustermodels.tab}
\begin{tabular}{r|r|r|r|r|r|r|r|r|r}\hline
Model & $\Gamma$ & $\sigma_{\mathrm{cl}}$ & $\overline{\sigma}_{\mathrm{v}}$ &
$\mathrm{var}_{\sigma}$ \\ \hline
A & 1.0 & 0 & 131.3 & 91.7\\ 
B & 0.9 & 558.5 & 118.2 & 82.6\\ 
C & 0.7 & 914.9 & 91.9 & 64.2\\ 
D & 0.5 & 1109.5 & 65.6 & 45.9\\ 
E & 0.2 & 1255.3 & 26.2 & 18.3\\
F & 0.0 & 1281.2 & 0.0 & 0.0\\ \hline
\end{tabular}
\caption{Parameters for 6 simulated mass models of clusters. The relative contribution of the individual galaxies, parameterised by $\Gamma$ is listed together with the central velocity dispersion of the smooth cluster halo. 
The Einstein radius is $32$\,arcsec for all models. The
tabulated values for $\sigma_{\mathrm{cl}}$, $\sigma_{\mathrm{v}}$ and
$\mathrm{var}_{\mathrm{\sigma}}$ are in $\mathrm{km\,s^{-1}}$.}
\end{table}
We set the size of the simulated strong lensing image to $2'\times2'$,
which corresponds  roughly to  the field of  view of the  Hubble Space
Telescope (HST).
The surface mass density maps of 6 different model clusters are shown in
Fig.\,\ref{clustermodelsA.fig} together with the simulated arcs (contours).
The different cluster models are summarised in Table\,1. Note that, for all models, the individual galaxy positions and redshifts remain fixed.

Our particular choice for the structure of the cluster model is simple. It is
motivated mainly by the requirement that the total mass distribution of the
cluster is close to the Navarro, Frenk and White profile
\citep[NFW,][]{navarro1997}. The galaxy and cluster halo mass distribution used
here gives a total mass profile that is nearly identical to an NFW profile.
 
In order to quantify the general lensing properties of the lensing cluster models we calculate magnification and surface mass density maps. These maps illustrate important general lensing properties of the cluster lens models. They are calculated using \emph{gLens} by mapping small triangles from image plane to source plane as described in \citet{moller1998}.
\begin{figure}
\epsfig{file=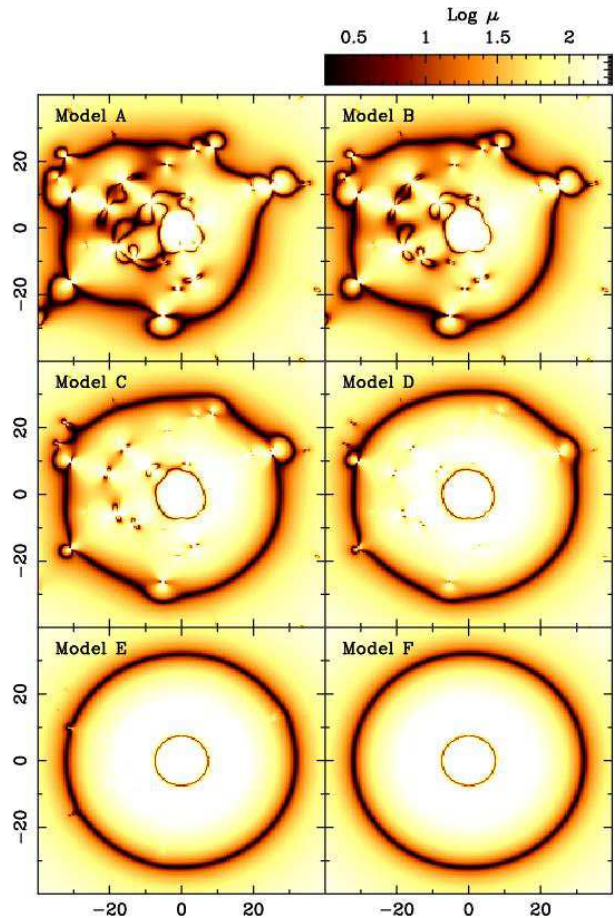,width=8.0cm,angle=180}\\
\caption{The image magnification maps of the model lensing
  clusters. The panels are for models A-F in Table\,1 and show
  the magnification of point sources as a function of image position.}
\label{clustermodelsB.fig}
\end{figure}
\begin{figure}
\epsfig{file=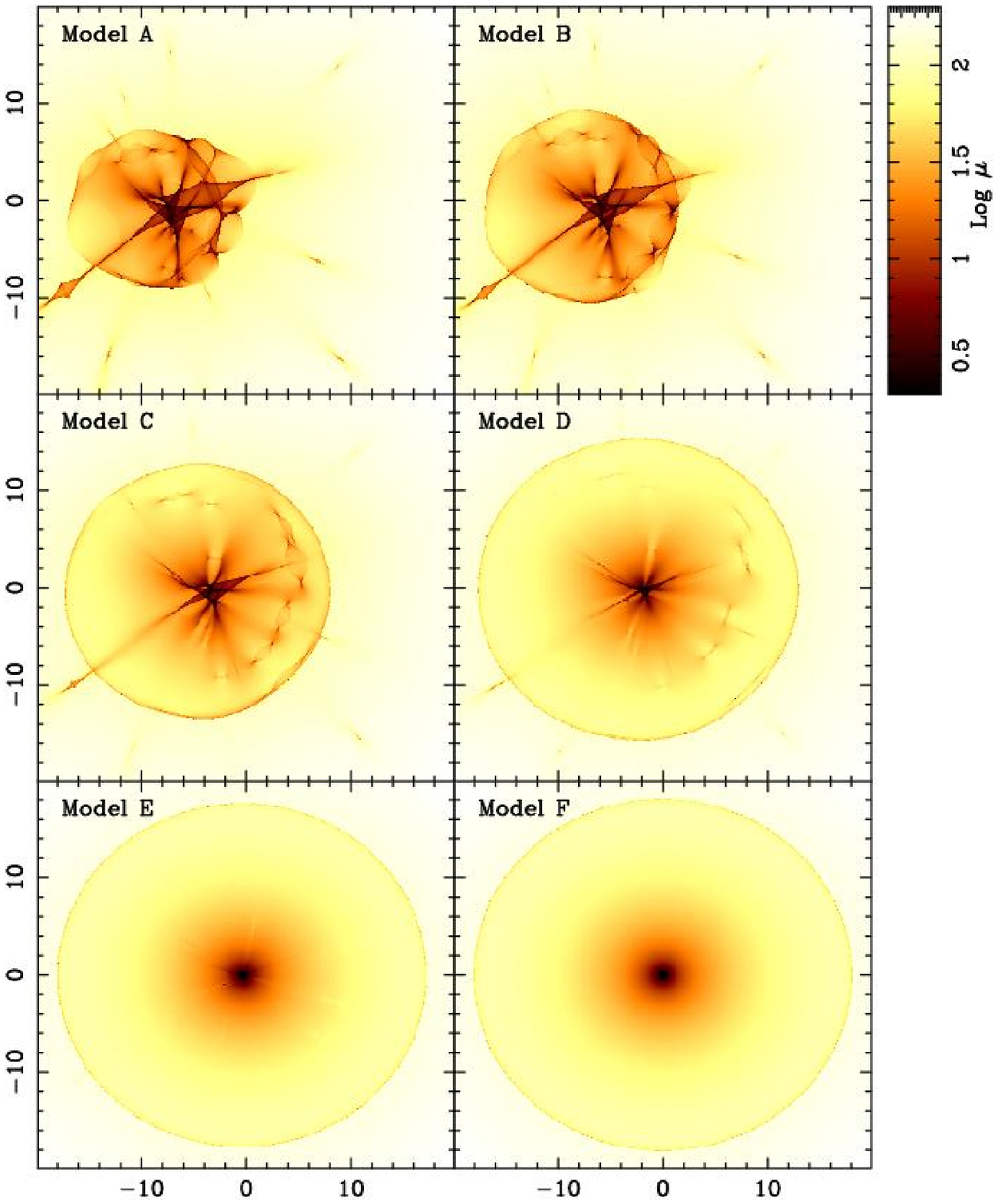,width=8.5cm,angle=180}\\
\caption{The magnification map on the source plane for the model
lensing clusters. As for Fig.\,\ref{clustermodelsA.fig}, the panels
are for models A-F in Table\,1. The scale is
logarithmic for all plots.  
}
\label{clustersourcemags.fig}
\end{figure}
Figs.\,\ref{clustermodelsA.fig} and \ref{clustermodelsB.fig} show the
surface mass density of the clusters we used as mass model template and 
the resulting magnification patterns on the image
plane. The magnification maps show the magnification $\mu$ of
a point source as a function of position. The corresponding magnification maps on the
source plane are shown in Fig. \ref{clustersourcemags.fig}. The high
magnification regions on the image and source planes are the `critical' and
`caustic' lines respectively. Note that the structure of the critical lines,
for $\Gamma>0.5$ are very similar to the ones obtained from lens mass
reconstructions of observed clusters \citep{kneib1996,broadhurst2005}.   

These figures demonstrate that increasing the mass in the individual galaxies, and
thereby increasing the amount of substructure, creates more strongly
distorted critical lines. Any long arcs of highly magnified
sources that are produced along the strongly curved sections of the critical lines will appear
broken and distorted. Therefore, the probability of observing
broken and distorted arcs increases with increasing fraction of the total
cluster mass that resides in individual galaxies. For most of the
remainder of this paper we will use a model with a very strong amount of
substructure (model A). This model probably represents an extreme case. Since
we expect the accuracy with which the source can be reconstructed to decrease
with increasing amount of substructure, this will be a `worst-case' scenario
for any reconstruction method.  

%
\subsection{Generating a lensed data cube using ray-tracing}
\label{lensedcube.subsec}
In  order  to obtain  simulated  images  of  the source  at  different
wavelengths, we  use the ray-tracing code \emph{gLens},  previously
used     for     several    lensing     studies
\citep{moller1998,moller2001,moller2003}. All pixels  of a given image
are mapped  from the image plane  onto the source plane.   The flux of
the $i$th pixel in the image plane, $f_i=F^{\mathrm{I}}(\theta_i)$, is set
using   the   flux  on   the   source   plane  $F^{\mathrm{S}}$   from
$F^{\mathrm{I}}(\theta_i)=F^{\mathrm{S}}\left(\theta_i-\alpha_i\frac{D_{\mathrm{LS}}}{D_{\mathrm{OS}}}\right)$. This
method works  well and is very  robust, but is not  the most efficient
way to calculate  images of extended sources. For  example, many pixels
are mapped  to the empty regions  in the source  plane. Ray-tracing of
these  `empty' rays  could  be  avoided using  some  sort of  adaptive
algorithm  \citep{moller2001}.  However,  for  the  purpose  here  the
non-adaptive  approach  is sufficient.  The  ray-tracing technique  is
extremely  accurate.  Numerical artefacts appear  only when  there is  a strong
mismatch  between  the   source  and   image  plane   resolutions  or
dimensions. In this paper, the image plane has a size of
$120\times120$\,arcsec  and  a  resolution  of  $N=1200\times1200$
pixels. This image plane is mapped onto  a source plane of  $800\times800$ pixel
resolution, covering an area  of $40\times 40$\,\arcsec. With
these settings numerical errors are negligible.
%
%
\section{Properties of lensed arcs}
\label{knownarcs.subsec}
Several   strongly  lensed   arcs   have  been   discovered  to   date
\citep[e.g.][]{fort1988,kneib1995,luppino1999,gladders2003,broadhurst2005}.   The arcs in Abell  2218 and
Abell 370  are perhaps the  most notable of  these, being up to  20\,arcsec long  and 2-3\,arcsec  wide.  Smaller
arcs have been discovered in  some other clusters, like CL 1358+62. In
these clusters, the  arcs are usually less extended  in both directions
with lengths of a few arcseconds and widths $\sim1$ arcsecond. It is important to make a
distinction here between  arcs that are produced by  galaxy lenses and
arcs that are  lensed by massive clusters. The  lengths and widths are
quite different.  An example of an arc produced in strong galaxy-galaxy
lensing is the most prominent arc in the Ultra Deep Field \citep[UDF]{blakeslee2004}. The lensing
galaxy is  a field elliptical, much  less massive than  the central cD
galaxies  in massive  clusters.  The  arc is  only $\sim0.3$\,arcsec wide and
about $\sim1$ arcsec long.

We  show  the  resulting images  of  our simulated
lensed    arcs as contours in
Fig.\,\ref{clustermodelsA.fig}. The  simulated arcs for
models A and B are very similar to those actually observed for several cluster
lenses and  show  the  main features clearly: a broken structure caused  by the individual galaxies in the
lens. The arc  lengths and  widths of $\sim10$\,arcsec  and $\sim2$\,arcsec,
respectively, are very comparable  to observed widths and lengths. The
smoother potentials  of models C-F  lead to continuous,  smoother arcs.
The appearance of the lensed arcs is very
similar to  observed arc systems for models A and B and  we therefore use  only these two
models for the remainder of this paper. 
\section{The velocity fields of simulated arcs}
\label{lensed.sec}
\subsection{Smooth sources}
Using      the       ray-tracing      procedure      described      in
\S\,\ref{lensedcube.subsec} we calculated  the individual channel maps
of      the  lensed    source,       modeled      as      described      in
\S\,\ref{lensmodel.subsec}. The velocity  fields of the strongly lensed
arcs  are determined  by fitting  gaussians to  the line  profiles. In this
section, we do not  include observational effects, like seeing, coarse
instrument  resolution   or  noise.    These  will  be   discussed  in
\S\,\ref{observations.sec}.  

\begin{figure*}
\epsfig{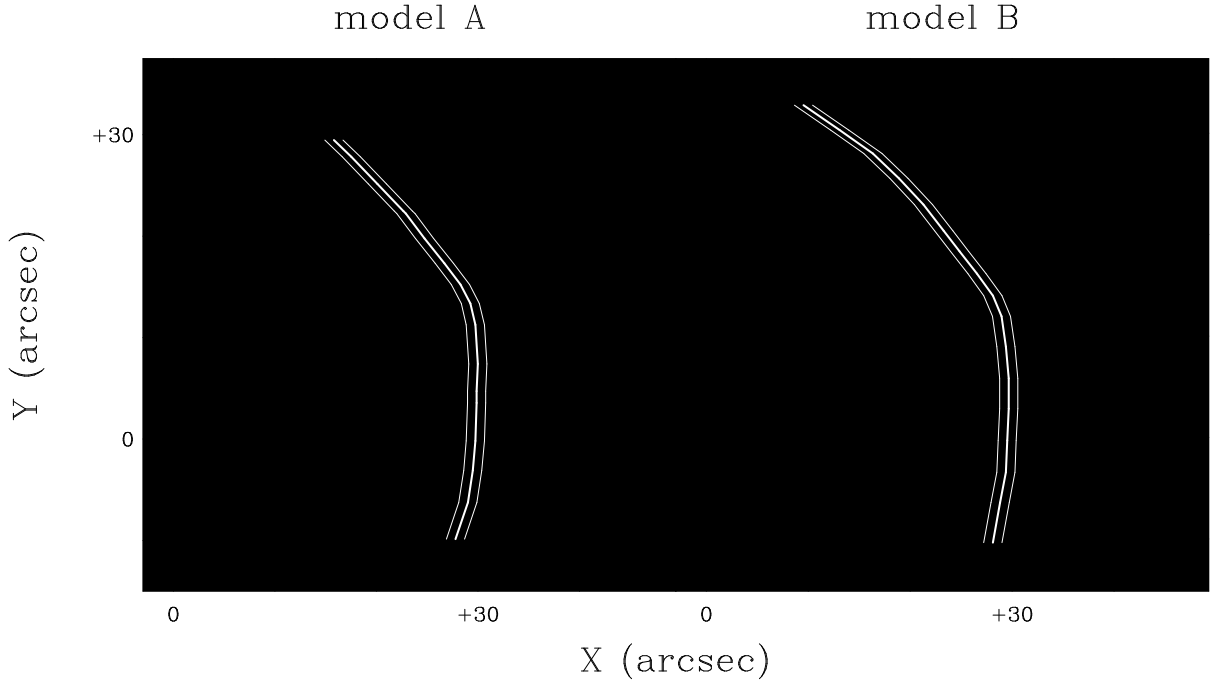}\\
\caption{Velocity fields for an extended  source lensed by a galaxy
  cluster. The parameters for the lens and source are as in the previous
  figure for model A and B. The white lines indicate slices along which the
  position-velocity diagrams are determined (shown in Fig.\,6). The colour scale is the same as for the inset in Fig.\,\ref{fig.single.vel}
}
\label{fig.cluster.channels}
\end{figure*}
We show the lensed velocity fields of the main arcs for models A and B in
Fig.\,\ref{fig.cluster.channels}. The resulting velocity structure is
very complex. Gravitational lensing into multiple images produces a
very distorted and asymmetric velocity structure along the arc. When
velocity information of a lensed arc is available, the source can be
essentially broken down into several smaller components which cover
different parts of the source plane and are therefore all magnified
and distorted in a different way. This effect of differential
magnification  was already discussed in \S\,\ref{theory.subsec} and,
in a different context, by \citet{blain1999}.
The small difference between models A and B in terms of the mass
distribution within the cluster (in model A, the galaxy that is
closest to the arc has a velocity dispersion of
$248\,\mathrm{km\,sec^{-1}}$, in model B it is
$223\,\mathrm{km\,sec^{-1}}$) translates into a noticeable and
measurable difference in the velocity fields: in model B, the regions
with approaching velocities are more strongly magnified relative to the
receding side.  
\begin{figure*}
\epsfig{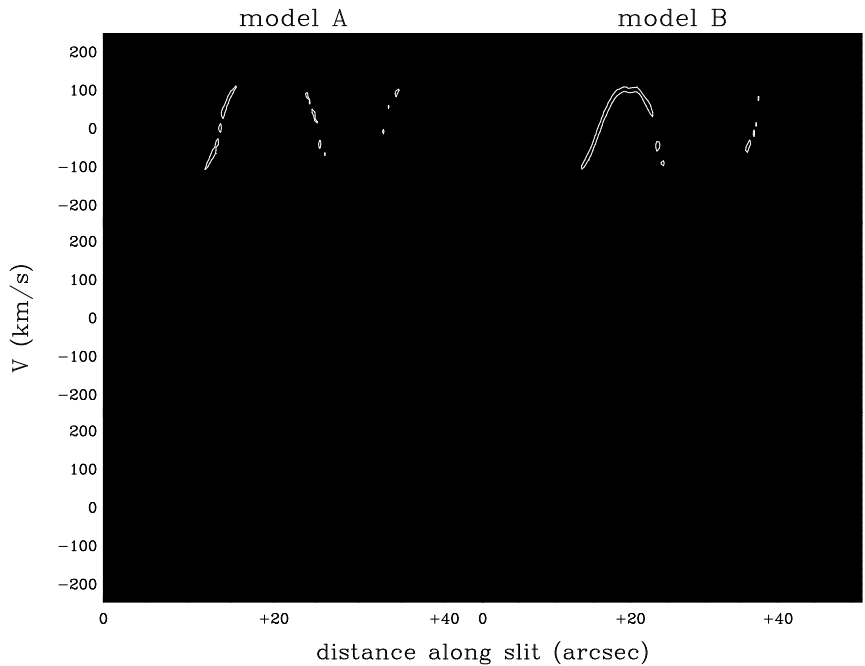}\\
\caption{Position-velocity diagram along the main lensed arcs. Only the
velocity profiles for models A and B are shown. Panels (a) and (b)
show the position-velocity diagram along a line following the
brightest pixels on the arc (the thick white line in Fig.\,\ref{fig.cluster.channels}). In panels (c) and (d) this line is offset
by +3 pixels ($\approx$ 1\,arcsec) in x, in panels (e) and (f) it is offset
by -3 pixels in x.   The contours correspond to the maximum
intensities at 1, 3 and 5 radial scalelengths in the input model 
(cf. Fig.\,\ref{fig.single.vel}). 
For all offsets the position-velocity diagram is heavily distorted
with respect to the unlensed case.  
}
\label{fig.cluster.vel}
\end{figure*}
The left and right panels in Fig.\,\ref{fig.cluster.vel} show the position
velocity diagrams for model A and model B, respectively. The positions of the
curved slits used to generate the position velocity diagrams are indicated in
Fig.\,\ref{fig.cluster.channels} by the white lines. They have a width of 1
pixel or $0.3\,$\,arcsec. A broader slit would increase the velocity range at
each position element and the curves would become broader.
The two lower sets of panels show the results if the original slit is
displaced by $\pm3$ pixels or $\pm1$\,arcsec. This changes the regions along the velocity
field of the lensed arc that are probed, leading to changes in the
shape of the curve. In particular, remarkable features like strong
asymmetry in the position-velocity diagram may result; as seen in
panels (c) and (d). 
Note that none of the lensed position-velocity slices
resemble those of unlensed galaxies. It is also noteworthy that the lensed
position-velocity slices shown here look very similar to those observed by
\citet[Fig.4]{pello1991}.

There is another simple and drastic effect that gravitational lensing has on
the velocity structure of sources, which can be demonstrated easily by looking
at the total flux emitted as a function of channel.
\afterpage{\begin{figure}
\epsfig{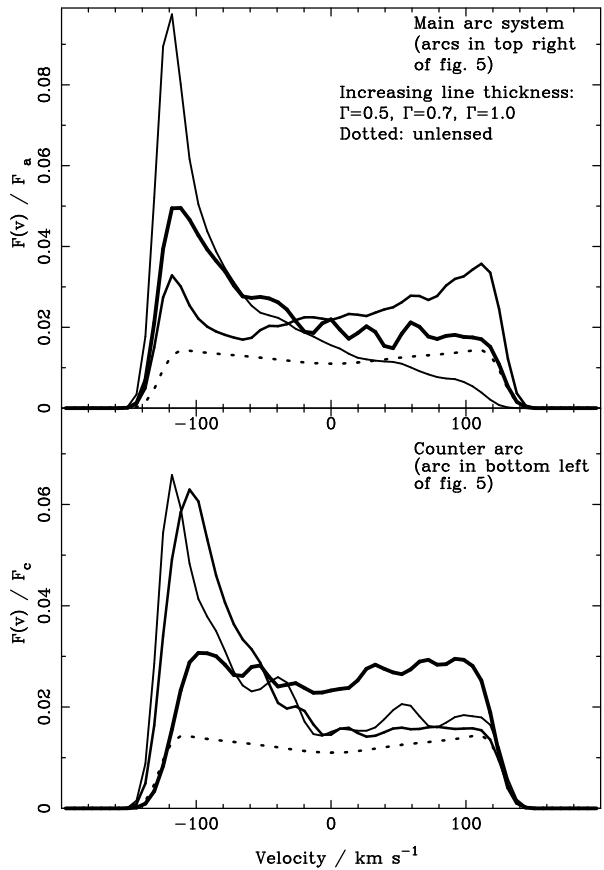}\\
\caption{The flux of the simulated image systems as a function of the
  channel. The ratio of the flux in each channel to the
  total flux in the arc is shown. The top panel shows the results for the main lensing arc
  and the bottom panel shows the results for the counter arc. With increasing
  line thickness, the curves are for $\Gamma=$0.5, 0.7 and 1. The dotted
  curves show the unlensed case. Note that, to increase clarity, the dotted
  curve is normalised differently than the solid curves.}
\label{fig.cluster.flux}
\end{figure}}
Fig.\,\ref{fig.cluster.flux} shows the flux ratio  
$F(\nu)/F_{\mathrm{tot}}$ in the main arc (top)
and the corresponding ratio $F(\nu)/F_{\mathrm{tot}}$
for the counter arc (bottom). In both cases, $F({\nu})$ and $F_{\mathrm{tot}}$
are given by eqs.\,9 and 7, respectively. Comparison with the unlensed case, as
shown by the dotted line in Fig.\,\ref{fig.cluster.flux}, shows the
effect of differential magnification very clearly. For most models, the
parts of the source with velocities around
$-100\,\mathrm{km}\,\mathrm{s}^{-1}$ are magnified more strongly than
the rest of the galaxy. For a symmetric galaxy such a profile is a strong
indication of differential magnification. For a single arc, this effect may
also be
reproduced without differential magnification when there is strong
asymmetrical substructure in the source \citep{richter1994}. Differential
magnification can either enhance or partially cancel intrinsic asymmetries in
the source. However, since the differential magnification is strongly
influenced by the small scale -- and hence local -- structure in the lensing
potential, lensing will in general produce different asymmetries in different
arcs of the same source. In the given case, a
comparison between the fluxes for the main arc, in the top panel of Fig.\,\ref{fig.cluster.flux}, and the
counter arc, in the bottom panel, shows that the asymmetries are induced by
lensing. In this way, asymmetries induced by lensing can be distinguished from intrinsic asymmetries, which affect all arcs of a given source. 
\subsection{Source substructure}
\label{subsource}
Throughout  we  have assumed  a  smooth  light  distribution for our
source galaxy. At high redshifts the merger rate is expected to be
much higher than in the local universe \citep{patton2002}, and
consequently  galaxies may  show far  more substructure  and kinematic
signatures  of  merger events  \citep{naab2003}.  Currently, there  is
still  some uncertainty  as  to how  much  of the  difference in  the
morphology and observed light distribution of high redshift galaxies is
due to intrinsic differences and how  much is due to the fact that the
optical at a redshift of $z\sim 1$ corresponds to restframe UV.  Due to
regions of star formation, local
late-type  galaxies are  found  to show  much  more substructure  when
observed in the UV.
\afterpage{\begin{figure}
\epsfig{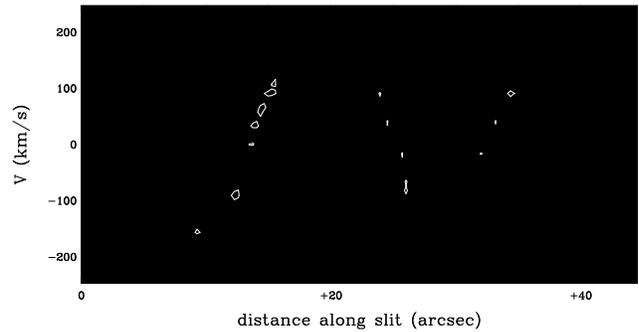}\\
\caption{Position-velocity diagram for a lensed clumpy source. The
source and lens model parameters are as for
Fig.\,\ref{fig.cluster.vel}, model A, except that the source now has a clumpy 
structure as indicated in the inset.} 
\label{fig.clumpy}
\end{figure}}
How  would our  results change  if  the source  light distribution  is
intrinsically less smooth\,?  As long as  the underlying
kinematics does not change, we  find that the clumpiness of the source
has  little   effect  on  the   results  discussed  in   the  previous
sections.  This is  illustrated in  Fig.\,\ref{fig.clumpy},  where we
show the  clumpy source  and the position  velocity diagram  along the
lensed arc,  which has the same  shape and position as  for the smooth
source model. Apart from many discontinuities in the diagram the shape
is  the same as  in Fig.\,\ref{fig.cluster.vel}.  However,  we still
assume dynamically  stable rotation  everywhere within the  source. In
particular our model  assumes that there are no  major in- or outflows
related to the source. If these are present, as may be expected for a 
fraction  of sources at high redshifts, these  will show up as
clear signatures in the  reconstructed velocity profiles of the lensed
arcs. We discuss in more detail how well substructured sources can
be reconstructed in \S\,5.5. 
\section{Reconstructing the source kinematics and the lens mass distribution}
\label{reconstruct.sec}
\afterpage{\begin{figure*}
\epsfig{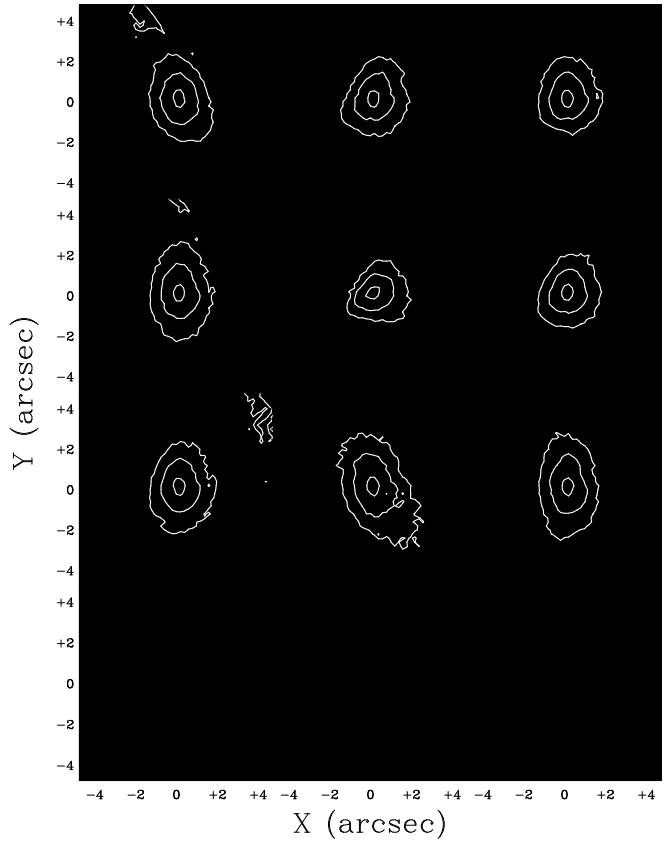}\\
\caption{The colour scales in the top 3 rows show the velocity fields of the 9 best-fitting reconstructions of the
smooth source, corresponding to lens model A. The bottom row shows the
velocity field of the three best reconstructions for the substructured source
described in \S\,\ref{subsource}. 
The lines show the isocontours of the corresponding images at 1, 3 and 5 scale lengths.} 
\label{fig.velbest}
\end{figure*}}

\subsection{Reconstruction algorithm}
In the previous sections we discussed the velocity structure of lensed arcs of extended sources. Using a simple model for the kinematic structure of the source, we have shown that the observed velocity structure in the arc is complex and depends on the mass distribution in the strong lensing cluster -- especially the mass in the galaxies close to the arc.

A remaining question is how well the original velocity structure of the source
can be reconstructed from the available channel information of the lensed
arc. Clusters of galaxies have been modeled from strong lensing in the past
\citep[e.g.][]{kneib1998}, deriving the cluster potential from the positions
and shapes of the arcs. This approach works very well when the overall cluster
potential is to be determined. When the source itself is to be reconstructed,
it is often necessary to perform more elaborate fits including pixel
information \citep{tyson1998}.

Using our simulated images for cluster model A, we attempt to reconstruct the
cluster and source model parameters from the image data alone. We assume that
galaxy redshifts and positions in the cluster are known, but not their
masses. A number of 14 galaxies are included in the cluster model, in addition
to a halo of unknown position and mass. The free source parameters are
position, exponential scale length, total flux, position angle and axis
ratio. In total we therefore have $14+3+6=23$ free parameters. We use a total
of $N_{\mathrm{pix}}=500$ pixels selected randomly from the region around the
arc on the image as constraints. Each pixel is mapped onto the source plane
and the flux at the source plane position is compared with the flux of the
source model. The positions of the pixels themselves also provide a
constraint: bright pixels should be clustered more than faint pixels. We
include this constraint by first calculating the flux weighed centre of the
mapped source pixels,
\begin{equation}
\mathbf{x_0}=\frac{\sum_{i=1}^{N_{\mathrm{pix}}}f_i\mathbf{x_i}}{\sum_{i=1}^{N_{\mathrm{pix}}}f_i}
\end{equation}
and then calculating the distance of each pixel with respect to this flux weighed centre.
The total `goodness', $\xi$ is then calculated as:
\begin{equation}
\xi^2=\sum_{i=1}^{N_{\mathrm{pix}}}|\mathbf{x_i}-\mathbf{x_0}|^2f_i^2+(f_i-m_i)^2.
\label{fit.eq}
\end{equation}
In this equation, $m_i$ is the model flux at pixel $i$. Note that we
include a dependence on the pixel positions $f_i$ in the first term, since in
our source model brighter pixels are required to be more compact than fainter
pixels. Also, note that this definition is only useful for determining the
best fit models -- a meaningful $\chi^2$ value can only be defined on the
image plane. This is done below in \S\,5.2.

In order to obtain an acceptable fit, and also to include possible
degeneracies, we perform the fitting using a modified simulated annealing
technique, with slow cooling. A population of 400 model clusters, initially
randomly sampling the parameter space in a uniform manner, is slowly adjusted
in 4000 steps. At each step, a new point in parameter space is chosen,
sampling the logarithmic parameter space using a Monte Carlo Markov Chain
(MCMC) method. After each step, the new set of model parameters replaces the old one with a probability given by
\begin{equation}
p_{\mathrm{replace}}=\mathrm{min}\left(e^{(\chi_{\mathrm{old}}-\chi_{\mathrm{new}})/T},1\right),
\end{equation}
where $T$ is the current `temperature' of the system, which is cooled from
$T_{\mathrm{start}}=10$ to $T_{\mathrm{end}}=0.001$ in the 4000 steps
logarithmically. From the final sample of 400 models, we select the 9
best-fitting models.
\afterpage{\begin{figure*}
\epsfig{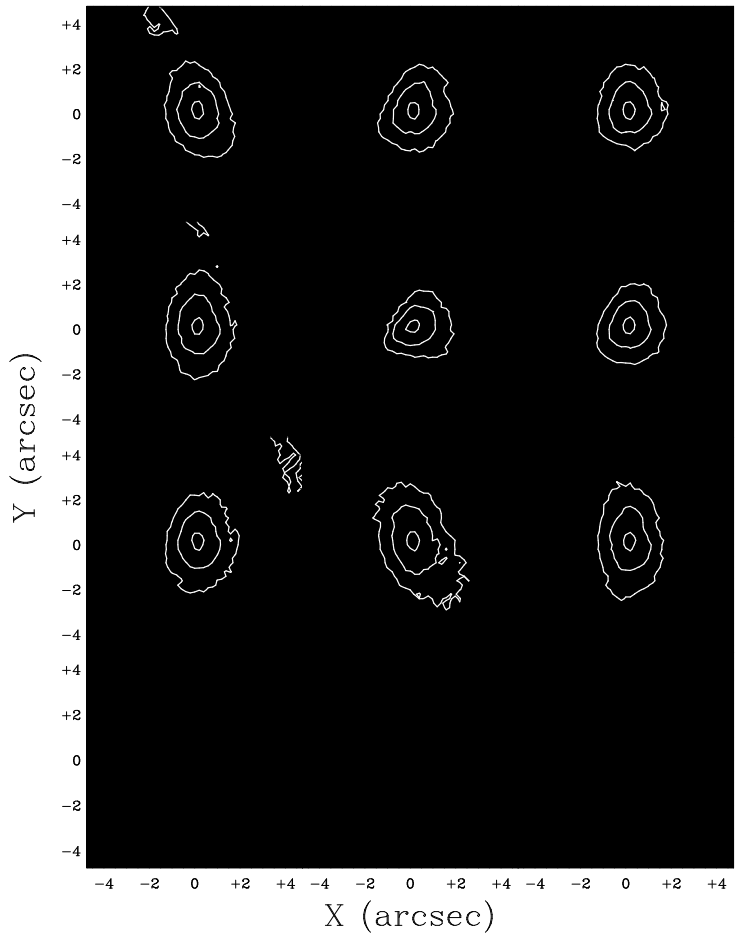}\\
\caption{Residuals between the reconstructed velocity fields and tilted ring model fits. The panels are for the same reconstructions as
in the previous figure. 
The lines show the isocontours of the reconstructed source image at 1, 3 and 5 scale lengths.} 
\label{fig.residual}
\end{figure*}}

\subsection{The reconstructed sources}
The contours in Fig.\,\ref{fig.velbest} show isophotes of the resulting source
reconstructions for the 9 best-fitting lens models. These source
reconstructions are calculated by ray-tracing the original,
`observed' image to the source plane through the corresponding
reconstructed model of the lens.  
All sources are compact and in rough agreement with the input position
angle and inclination. Most importantly all the best-fitting models
give source models with an integrated light profile that is consistent
with an exponential disc profile. 

We define a goodness of fit in the image
plane as
\begin{equation}
\chi^2=\sum_{i=1}^{N}\left(\frac{f_i-m_i}{\sigma_i}\right)^2w_i,
\end{equation}
where 
\begin{equation}
w_i=\frac{f_i+m_i}{\sum_{j=1}^{N}f_j+m_j},
\end{equation}
and $\sigma_i=5\times10^{-5}$ is our assumed surface brightness error in units
of the total flux of the source. 
With this definition, $\chi^2\sim1-3$ in all of these cases.
The velocity fields of the best-fitting reconstructed sources are
displayed with the colour scales in Fig.\,\ref{fig.velbest}. Most of
the reconstructed velocity fields show the same global shape as the
input model, but in several cases distortions are present, especially
in the outer parts. These distortions are inconsistent with
dynamically stable rotation and can be used to distinguish between the
different reconstructions.  

To investigate which reconstructed velocity fields are consistent with
regularly rotating gas discs, we tried to fit each of them with tilted
ring models. In these fits, gas is assumed to move on circular orbits
around the centre of the galaxy in a series of concentric rings. The
position angle and inclination of the galaxy, as well as the rotation
velocity of each ring is fitted to obtain the best match with our
`observed' reconstructed velocity field. The  differences between
the lens reconstructions of the source kinematics and our best tilted
ring fits are shown in Fig.\,\ref{fig.residual} for all
reconstructions. For the majority of cases, the differences in the central
regions of the source are small, but larger in the outer, high-velocity
regions of the source. However, several reconstructions (e.g. 2, 5 and 8) show
significant distortions in the inner regions. Such distortions are not
observed in real galaxies and must therefore be due to an imperfect lens model. \subsection{Comparison of isophotal and kinematic fits}
Even though  the source  light  profile is  acceptable for  all
reconstructions, close inspection  of the velocity fields  shows that
some of the  lens models are insufficiently accurate to allow
reconstruction of the source kinematics. To  investigate this further, we also
performed  an isophotal  analysis of the reconstructed images and
compared the morphological orientation of the reconstructed sources
with the kinematical orientation as derived from the tilted ring fits.
\begin{table}
\label{sourcemodels.tab}
\begin{tabular}{r|r|r|r|r|r|r|r|r|r}\hline
Reconstruction & $\phi_{\mathrm{kin}}$ & $\theta_{\mathrm{kin}}$ &
$\phi_{\mathrm{iso}}$ & $\theta_{\mathrm{iso}}$ \\ \hline
1 & 6$\pm$2 & 39$\pm$5 & 14$\pm$2 & 51$\pm$1\\
2 & -1$\pm$1 & 33$\pm$6 & -16$\pm$3 & 40$\pm$2\\
3 & 1$\pm$1 & 37$\pm$4 & -1$\pm$1 & 39$\pm$2\\
4 & 1$\pm$1 & 49$\pm$2 & 1$\pm$1 & 51$\pm$1\\   
5 & 13$\pm$2 & 25$\pm$10 & -59$\pm$3 & 35$\pm$3\\   
6 & -2$\pm$1 & 34$\pm$4 & -14$\pm$1 & 39$\pm$2\\
7 & -2$\pm$1 & 44$\pm$2 & 0$\pm$2 & 46$\pm$1\\
8 & 13$\pm$2 & 42$\pm$5 & 17$\pm$2 & 55$\pm$1\\  
9 & -2$\pm$1 & 52$\pm$1 & 3$\pm$ 1 & 55$\pm$1\\
C1&  0$\pm$1 & 40$\pm$4 & 11$\pm$7 & 39$\pm$4\\  
C2& -3$\pm$2 & 33$\pm$5 & 2$\pm$12 & 38$\pm$4\\
C3& -4$\pm$1 & 51$\pm$2 &-21$\pm$2 & 49$\pm$3\\ \hline
\end{tabular}
\caption{
Parameters of the source reconstructions for the 9 best-fitting lens
models. The values for the position angle from isophotal and tilted
ring fits, $\phi_{\mathrm{iso}}$ and $\phi_{\mathrm{kin}}$ are listed
together with the corresponding inclination angles, $\theta_{\mathrm{iso}}$
and $\theta_{\mathrm{kin}}$. Also shown are the results for the substructured
models C1, C2 and C3, discussed in \S\,5.5.} 
\end{table} 
The derived values for the position angle and inclination from both
analyses are listed in Table\,2. 
Comparing the position angles of  the isophotal analysis to  those
obtained  from  the   tilted  ring  fits shows a large discrepancy
of 10 or more degrees  for several reconstructions  (2, 5, 6). The
inclination angles from the isophotal and  tilted ring fits agree to
within 10 degrees in all cases. The inclination angles from isophotal fits 
are within 10 degrees of the input
value of 50 degrees with  the  only  exception  of  reconstruction  2
which has a  best  fit inclination angle of 35  degrees. Only for
models 3, 4, 7 and 9 is
$|\theta_{\mathrm{iso}}-\theta_{\mathrm{kin}}|\lesssim5$ 
degrees. Inspecting the residuals between the 
velocity  fields  as  fitted  with the tilted  ring  models  and  the
reconstructed  velocity fields, shown  in Fig.\,\ref{fig.residual},
also helps to discriminate different models.  The strongest residuals
in the central parts -- that is inside the contours in each panel of
Fig.\,\ref{fig.residual} -- are associated with the 
reconstructions for  models 1, 2, 5 and 8. 

On the basis of the information from Table\,2   
and Fig.\,\ref{fig.residual}, one would draw the conclusion that
reconstructions 3, 4, 7 and 9 are the most accurate. From these, one
would constrain the position angle and inclination angle of the source
to be $\frac{1}{2}(\phi_{\mathrm{kin}}+\phi_{\mathrm{iso}})=0.2\pm0.5$\,degrees
and $\frac{1}{2}(\theta_{\mathrm{kin}}+\theta_{\mathrm{iso}})=47\pm3$\,degrees --
consistent with the input values.
\begin{figure}
\epsfig{file=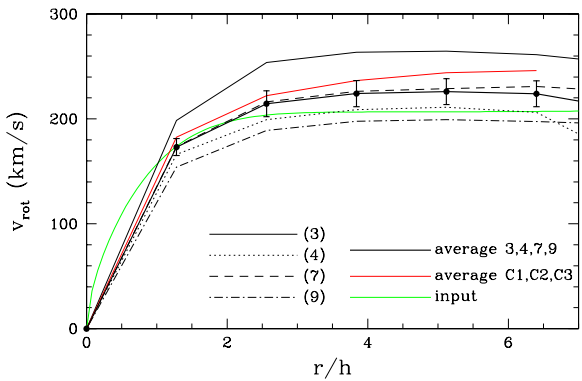,width=8.0cm,angle=0}\\
\caption{Rotation curves from tilted ring fits. The thin black lines
give the rotation curves derived from the velocity fields from
reconstructions 3, 4, 7 and 9 from Table\,2. The
bold line with data points shows the average of the 4 individual
curves; the errorbars are estimated as 1/2 of the rms scatter of the
points from the individual curves. The solid, thin green line shows the input
rotation curve from eq.\,\ref{URC.eq}. The red line shows the result for the
average of the three best fits to the substructured source (cf. \S\,\ref{subsource}).} 
\label{fig.rotcurs}
\end{figure}
In Fig.\,\ref{fig.rotcurs}, we show the rotation curves derived from
the tilted ring fits for these reconstructions, together with the
average of all 4 and the input rotation curve from
eq.\,\ref{URC.eq}. Within the expected observational
uncertainties these reconstructed velocity
fields allow an accurate recovery of the input rotation curve. 

\subsection{Effect of cluster mass profile and ellipticity}
\label{nfwsect}
Thus far we have assumed a cored pseudo-isothermal model for the cluster halo mass
distribution in our model. The
\emph{total} mass profile in our model cluster follows an NFW form only due
to added substructure in the form of galaxies.  A pseudo-isothermal mass profile has been shown to provide good fits to lensing
cluster halos in previous studies \citep{sand2005,kneib1996}. Numerical simulations, however,
predict that cluster halo mass profiles follow a NFW form more closely. In addition, we have assumed a spherical mass
distribution for the cluster, whereas most real and simulated clusters have
elliptical mass profiles. There is a
degeneracy here between the ellipticity of the halo mass profile and the presence
of massive substructure, as both induce an
asymmetric lensing potential. The effect of triaxiality on lensing statistics
has also been found to be degenerate with core size by \citet{oguri2005}. We have tested how strongly ellipticity in the
lens model affects the properties of the arcs and the reconstructed
source. 
\begin{figure}
\epsfig{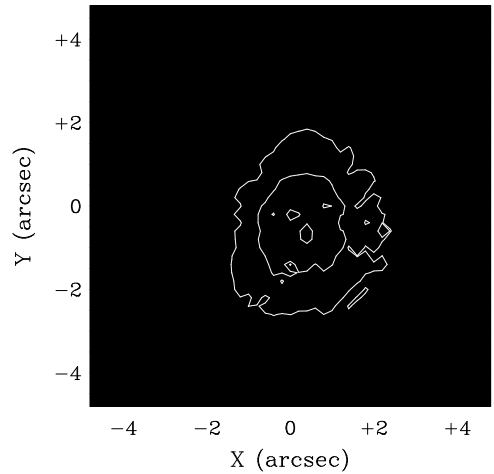}\\
\caption{The reconstructed source kinematics of the best-fit elliptical NFW cluster lens
  model. The original
  simulated data (shown in Fig.\,\ref{clustermodelsA.fig}) is fitted with an elliptical NFW halo
  model with three cluster galaxies. The figure shows the reconstructed
  integrated light profile (contours) and velocity field (colour) of the source.  The
  best-fit NFW lens model has an
  ellipticity of $\epsilon=\sqrt{1-b^2/a^2}=0.15$, where $a$ and $b$ are the major
  and minor axes, respectively. Its virial mass $M_{\mathrm{vir}}=1.925\times10^{14}M_{\odot}$ and its concentration parameter is $c=5.6$ (see \citet{navarro1997} for a definition
  of these parameters).
} 
\label{nfw.fig}
\end{figure}
Fig. \,\ref{nfw.fig} shows the reconstructed
source properties for the best fitting model of an elliptical NFW cluster
halo with three additional cluster galaxies. A relatively good fit to the
source light profile is retained. However, the
model fails in reproducing the small scale details of the velocity structure of the
source; the velocity field in the central region is bend into an `S'-shaped
structure. In principle, high quality kinematic
data could break the degeneracy between halo ellipticity and massive
substructure. In practice, however, it is doubtful whether current instruments
could provide data that would be accurate enough to detect these small scale
differences.
  
\subsection{Fits to substructured sources}
Above we discussed how the appearance of the velocity field of arcs would change if
sources are substructured instead of smooth. Since such substructures would
produce additional uncertainties in any source reconstructions, we also perform
fits to a substructured source model using our method.
The velocity fields and residuals to tilted ring fits for the three best
fitting reconstructions C1-C3 are shown in the bottom row of
Figs.\,\ref{fig.velbest}\,\&\,\ref{fig.residual}, respectively. Models C1
and C2 show a low residual to the tilted ring fit, whereas model C3 shows
some velocity distortions in the central region and appears slightly more
compact overall. Averaging the resulting best-fit rotation curves for these three
reconstruction gives an averaged inclination angle of $42\pm4$deg and an average
position angle of $4\pm1$ deg. These values are within 2$\sigma$ of the input. The averaged rotation curve is shown as the red line in
Fig.\,\ref{fig.rotcurs}.
\subsection{The lens mass reconstructions}
We investigated what differences in the lens mass reconstructions lead
to   the  observed differences  in   source  reconstructions.   Inspecting  the
reconstructed  mass maps  and calculating  the total  mass  inside the
Einstein radius for  all the reconstructions, we found  that there are
only very small differences between the different reconstructions; the 
total mass within an Einstein radius is always within 1-3\% of the input
value. However,  there  are  larger  differences in  how  the  mass  is
distributed within  the Einstein  ring; the reconstructions  differ in
their amount  and direction  of asymmetry in  the central  regions. In
addition, the surface mass density in regions away  from the main  arc vary  by $\sim30-50\%$
between  the reconstructions.  This large variation is  to be  expected  as the  main
constraints on the  lensing mass distribution comes from  the main arc
itself and  is, strictly, only  a \emph{local} constraint on  the mass
distribution. It  is only due to  the parameterised form  of the input
model that the arc constrains other  regions in the cluster at all. 
The  differences  in
the reconstructions of the velocity fields  mainly come from  the change  in the  distribution of
mass in the  central part of the cluster and from  small changes in the
mass of the  galaxies very close to the  main arc.
\section{Observational possibilities}
\label{observations.sec}
In  the previous  sections, we  have shown  how  gravitational lensing
affects the velocity fields  of high-redshift background galaxies, and
how the  additional information contained in  the observed
velocity  fields can be used to  constrain models  of the lensing cluster and
source galaxy.  In this 
section, we  will discuss  the possibilities of  observing the
velocity fields of strongly lensed galaxies.

It  is clear  from the  previous sections  that one  needs  to measure
velocities over the  full 2D extent of an arc, in order to use the
kinematic information to constrain the lens model and to determine the
rotation curve of the source galaxy. 
Simple long slit spectra along lensed arcs lack information about the
orientation of the source galaxy, making it impossible to
interpret the kinematic properties of the arc
unambiguously. Additionally, small slit  offsets and finite slit
widths  have a strong effect on  the observed profiles, as can be
appreciated from the difference between for example panels (a), (c)
and (e) in Fig.\,\ref{fig.cluster.vel}. 
 
To observe the velocity fields of giant arcs, integral field
spectroscopy  at high  spatial resolution  and high  sensitivity is
required.  Our simulated  lensed  velocity  maps in
Fig.\,\ref{fig.cluster.channels} have a width of about $5$ arcseconds,
but one should bear in mind that in producing these  figures, we have
not applied any  flux-cut. In reality, velocities  can only be
measured in the brighter  regions of the arcs, with a typical width of
$1$ arcsecond. Spatial resolution of at  least 0.2--0.3 arcseconds  is
required  to  resolve these regions. The demands on spectral
resolution are less stringent. Typical spiral galaxies have rotation
velocities in the range 100--300 km/s, so a velocity resolution of
order 50--100 km/s is sufficient to measure radial velocities  of the
gas to  a fraction of the expected rotation velocities.

To carry out the required observations, a number of options are
available.  Currently, the best opportunity is offered  at optical or
near-infrared  wavelengths, where, depending on  the  source
redshift,  several  strong emission  lines  of 
ionised  gas (e.g. H$\alpha$, O{\sc ii}, O{\sc iii}, etc.)  are
available. Sub-arcsecond seeing  is now  routinely achieved 
with adaptive  optics systems at a number  of ground-based telescopes,
and the number of integral field spectrographs that make use of the
high resolution offered by these systems is rapidly
increasing (e.g. GMOS on Gemini and Sinfoni on the VLT). 
The biggest obstacle currently seems to be that, in order for adaptive
optics systems to deliver the sub-arcsecond images, a bright ($m_V \lesssim
14$) guidestar  in the immediate neighbourhood  of the object is
required. Since most lensed arcs do  not lie close  enough to
such a bright  star, one  has  to await  the  development of 
artificial laser guide-star  systems to  observe the  most interesting 
arcs. However, technology seems to be improving rapidly, and several
observatories expect a working system within a few years from now.

All  giants  arcs  observed  hitherto  are intrinsically faint, so
large  telescopes  are required  to  obtain useful spectra.  Long-slit
spectra  of a  straight arc  at $z=0.91$ have been obtained by
\citet{pello1991},  using a $2$\,arcsec wide slit at the 4.2m  WHT. In
15  hours of  integration  time, they  obtained a  high
signal-to-noise   ratio  spectrum  which   enabled  them   to  extract 
velocities along  the full  length ($> 10$\,arcsec)  of the
arc. Several other groups have recently measured spatially resolved
velocities in unlensed galaxies out to redshifts of $z \sim 2$, using
8--10m class telescopes like Keck or VLT and slitwidths of 
0.5--1.0\,arcsec \citep{vogt1996,vogt1997,boehm2004,erb2004}. These
results imply that at 8--10m  class telescope like the VLT, Keck or 
Gemini,  sub-arcsecond resolution  observations should  be  feasible in
1--2 nights of integration time. 

Other  options lie  further in  the  future. ALMA  is currently  being
constructed, and will offer the required resolution and sensitivity to
observe the kinematics of molecular gas at the redshifts of the arcs
we study here. 
Even further ahead, giant radiotelescopes like SKA will  be able  to
observe  the H{\sc i} emission line  of neutral hydrogen. This  would
offer  the fascinating possibility  of measuring the kinematics  of
lensed galaxies  well outside their  stellar discs, probing  into  the
dark  matter  dominated  regions  of  these  young galaxies.
Finally, several studies are currently underway to design the next
generation optical telescopes, with diameters of 25 meter and
larger. With the light gathering powers of such extremely large
telescopes, it will be feasible to detect emission lines out to large
galactocentric distances in lensed arcs within very short exposure
times, thus enabling systematic studies of the kinematics of these
high-redshift galaxies.  
\section{Discussion and Conclusions}
\label{conclusions.sec}
Determining the properties of high redshift galaxies remains one of the main goals of current research. In this paper, we presented a first theoretical investigation on how the effect of gravitational lensing can be exploited to determine the kinematic properties of high redshift galaxies.

Using a parametric cluster model we simulated the velocity structure of strongly lensed background galaxies.  The combination of ray-tracing techniques with parametric cluster and source models proved to be a very efficient and accurate method for this study. 

In general, we found that the two dimensional kinematic profile along strongly
lensed arcs is very complex. Differential magnification leads to very
distorted position-velocity diagrams and strong asymmetries in the velocity
fields. Here, it is important to note that the individual cluster galaxies
close to the arc contribute strongly to this effect, as we demonstrated in
sections 4 and 5. Using a relatively simple-minded technique, we showed that
reconstructions of the 2D kinematic source properties of lensed arcs are in principle possible. Since the velocity structure is sensitive to small variations in the lensing potential, kinematic information along the arc provides additional tight constraints on the mass distribution in the proximity of lensed arcs. Observationally, the use of an integral field spectrograph at an 8--10m class telescope
with sub-arcsecond seeing will allow accurate source reconstructions and
measurements of the rotation curve of strongly lensed arcs. 
We predict that the inclination and position angles of sources that are dynamically stable
rotators can be determined to an accuracy of $\sim10\%$. Rotation curves can thereby be determined with accuracies of better than $15\%$ out to 2--3 disc scale lengths for galaxies at redshifts above $z\sim1.5$ in this way.

Our general approach was motivated by our aim to provide a general discussion of the kinematic properties of lensed arcs and point out the uses and possibilities of kinematic data of such systems. The parametric cluster model we used was simple, but it reproduced the observed appearance of strongly lensed arcs well, when the galaxy mass-fraction inside the cluster Einstein radius was high. We note here that it should
be possible to constrain the mass fraction of galaxies in clusters by
making statistical predictions about the shape of arcs, for example from N-body simulations,
and comparing them with the observed structure of arcs. A thorough study of this would have to take into account the effect of cluster merging and the mass function of cluster sub-haloes. 
Our predictions for the general appearance
of velocity fields of arcs are independent of the specific form of the cluster
potential, as long as the observed properties of arcs are reproduced. 
This is because any other description of the cluster potential must also reproduce the
observed properties of lensed arcs.
In particular, the local differential magnification that produces the distortions in the velocity fields arises from small scale mass structures close to the arcs, which are also responsible for the broken structure of observed arcs.

Our method here has made use of a smooth source model. Even though we
demonstrated that structure in the source does not change our results and does
not affect the appearance of arcs and their velocity fields significantly,
this is strictly only true as long as the source itself has stable
rotation. For high redshift sources that will probably not always be true, since
mergers, outflows etc. are much more common at higher redshifts. However, as
we pointed out, one of the advantages of the multiple arc systems formed by
lensing is that intrinsic properties of the source can be disentangled from
lensing induced distortions. Lensing distortions will be different for each
arc, since the local cluster mass structure is important, whereas intrinsic
source properties are the same for all arcs from a single source. In fact,
this can be exploited to the extent that the source and lens can be
reconstructed from arcs without (almost) any assumptions about the source
itself. \citet{warren2003} describe a method that exploits this and can
reconstruct the lens and source in the presence of noise and finite
seeing. Such a non-parametric method was previously described by
\citet{wallington1996,wallington1994} and extended by
\citet{koopmans2005}. It can be applied equally well to reconstruct
the kinematic source properties, independent of any assumptions about the
source -- except that the source be of a physically plausible size. 

In this paper we concentrated entirely on arcs produced by clusters. This was
motivated by the fact that cluster arcs are larger and easier to distinguish
from light emission originating from the lens plane. Galaxy lenses produce
considerably smaller arcs that are superimposed on the lens galaxy itself. However,
with IFU's of high spatial resolution it may be possible to determine the
kinematic profile of arcs lensed by galaxies as well. Since the relative scale
of source to lens is about unity for galaxy lenses, in contrast to cluster
lenses where it is much smaller, the appearance of the arcs produced by galaxy
lenses is much smoother than for cluster lenses. This can be explained by
noticing that a version of, for example, the top left panel in
Fig.\,\ref{clustersourcemags.fig} that is scaled down by a factor of 10 or
more would be covered almost completely by the source. This means that small
scale structure in the lensing potential would have almost no effect on the
overall appearance of the lensed arcs. However, if kinematic data is available, the
situation changes. Differential magnification becomes important for each
individual channel since a given velocity channel only probes a very small region
on the source plane. In other words, kinematic data of arcs behind galaxy lenses
can provide strong constraints on the amount of mass in small scale structures
of galaxy haloes, and may be used as a direct probe of the mass function at
the low-mass end -- possibly down to $\sim10^{7}M_{\odot}$. We will address
this in more detail in a future publication.

In summary, it is clear that obtaining two dimensional kinematic profiles of strongly lensed arcs will provide very useful information of both source and lens. Using IFU's in the very near future to study these systems will provide a unique way to
determine the rotation curve of strongly magnified galaxies at redshifts $z\sim 1.5$ or higher and measure their mass-to-light ratio out to several disc scale-lengths.
\section*{Acknowledgements}
We would like to thank Leon Koopmans, Simon White, Ben Panter and an anonymous
referee for useful comments on the manuscript. OM gratefully acknowledges
financial support from the Marie Curie Fellowship programme of the European Union.  
\bibliographystyle{mn2e}
\bibliography{mn-jour,Bibs}

\end{document}